# CONATION: English Command Input/Output System for Computers

**Kamlesh Sharma\* and Dr. T. V. Prasad\*\***
\* Research Scholar, \*\* Professor & Head
Dept. of Comp. Sc. & Engg., Lingaya's University, Faridabad, India
Email: kamlesh0581@gmail.com, tvprasad2002@yahoo.com

*Abstract* – **In this information technology age, a convenient and user friendly interface is required to operate the computer system on very fast rate. In human being, speech being a natural mode of communication has potential to being a fast and convenient mode of interaction with computer. Speech recognition will play an important role in taking technology to them. It is the need of this era to access the information with in seconds. This paper describes the design and development of speaker independent and English command interpreted system for computer. HMM model is used to represent the phoneme like speech commands. Experiments have been done on real world data and system has been trained in normal condition for real world subject.**

## 1. INTRODUCTION

The primarily human computer interaction is done by keyboard and mouse as pointing device work as input and monitor and printer work as output. Keyboard, although a popular medium is not very convenient as it requires a certain amount of skill for effective usage. A mouse on the other hand requires a good hand-eye co-ordination. It is also cumbersome for entering non-trivial amount of text data and hence requires use of an additional media such as keyboard. Physically challenged people find computers difficult to use. Partially blind people find reading from a monitor difficult.

With the integration of computers and telecommunications, the mode of information access becomes an important issue. The designs of the prevalent human machine interfaces are more suitable for easier interpretation of information by computers than by human beings. The concept of machine being able to interact with people in a mode that is natural as well as convenient for human beings is very appealing. Issuing spoken commands to a machine to get useful work done and to get the response is a no long dream now. This has motivated research in speech recognition as well as speech synthesis. Considerable progress has been made and a few commercial speech products of varying capabilities are available for use in quite a few languages.

English command Input/Output system for computer is a fascinating field spanning several areas of computer science and mathematics. Reliable speech recognition is a hard problem, requiring a combination of many techniques, however modern methods have been able to achieve an impressive degree of accuracy. This project attempts to examine those techniques, and to apply them to build a simple system. These are exciting technologies that change the way to interact with computer. To talk the computer using a set of pre-define commands and instruction and computer will respond in the same way. For example to say: "file open", and the computer will open a new file: "select the file". or "Edit find" and the computer will do all this work according to the word spoken to the system. The intent in developing this project is to ability to command and control the computer through voice. Speech recognition is a technology that allows the computer to identify and understand words spoken by a person using a microphone. [4][5]

The ultimate goal of this paper is to outline a system that can recognize all words that are spoken by any person and perform corresponding command. Computer software that understands the speech and conversation with the computer. This conversation would include person and computer, speaking as commands or in response to events, input or other feedback. Speaking is easier and more intuitive than selecting buttons and menu item, human speech has evolved over many thousand of year to become an efficient method of sharing information and giving instruction. The dynamic nature of the world only emphasized this need strongly.[1]

The system recognizes voice of any individual, records it, and matches with the respective command available and performs the action required. Database connected by commands to execute the function, wish to operate. Speech recognition systems have been around for over twenty years, but the early systems were very expensive and required powerful computers to run. The technology behind speech output has also change. Early system used discrete speech, i.e. the user had to speak one word at a time, with short pause between the words.

## 2. RELATED WORK

Dragon Dictate is the only discrete speech system still available commercially. Over the past few years most systems have used continuous speech, allowing the user to speak in a more natural way. The main continuous speech systems currently available for the PC are Dragon Naturally Speaking and IBM Via Voice. Microsoft has included their own speech recognition system within recent versions of Windows. There is now a version of IBM Via Voice for recent Apple MAC Computer. The aim of speaker recognition is to recognize the speaker while speech recognition is related to the detection of speech. Speech recognition has been a goal of research for more than four decades. [6][7]

## 3. COMMANDS RECOGNITION USING HMM

We need to recognise a word using the existing models of words that we have. Sound recorder need to record the sound when it detects the presence of a word. This recorded sound is then passed through feature vector extractor model. The output of the above module is a list of features taken every 10 msec. These features are then passed to the recognition module for recognition. The feature vectors generated by the feature vector generator module act as the list of observation for the recognition module. Probability of generation of the observation given a model, $P(O/\lambda)$, is calculated for each of the model using find probability function. The word corresponding to the HMM, that gives the probability that is highest and is above the threshold, is considered to be spoken.

*Forward Variable*
Forward variable was used to find the probability of list of occurrence given a HMM. For a model ˏ with *N* states, $P(O/\lambda)$ probability of observation, in terms of forward variable α, given the model is define as

$$P(O/\lambda) = \sum_{i=1}^{N} \alpha_T(i)$$

where $\alpha_{(t+1)}$ is recursive defined as

$$\alpha_{(t+1)} = [\sum^{N} \alpha_T(i) a_{ij}] b_j(O_{t+1})$$

i=1

where $\alpha_1$ is $\pi_i b_i(O_1)$

*Occurrence Probability*
For the forward variable to work we need to find $b_i(O_t)$. This is probability of a given occurrence for a particular state. This value can be calculated by Multivariate normal distribution formula. Probability of observation X occurring in state i is given as:

$$(1/(2\pi)^{D/2}|V_i|) \exp(-(1/2)*(O_t-\mu_i)^T V^{-1}(O_t-\mu_i))$$

where D is dimension of the vector,
$\mu_i$ is matrix representing the mean vector,
$V_i$ is the covariance matrix,
$|V_i|$ is the determinant of matrix $V_i$,
$V^{-1}$ is the inverse of matrix $V_i$.

Mean vector $\mu_i$ is obtained by:

$$\mu_i = (1/N)* \sum O_t \; O_t \epsilon i$$

**Covariance Matrix Vi can be obtained by:**

$$V_i = (1/N)* \sum (O_t - \mu_i)^T * (O_t - \mu_i) \; O_t \epsilon i$$

Variance is calculated by finding the distance vector between an observation and the mean. Transpose of the distance vector is taken and it is multiplied with the distance vector. This operation gives a N x N where N is the dimension of the system. [2]

## 4. TRAINED THE SYSTEM

To train the system we required three parameters:
- No of states the HMM model should have N.
- The size of the feature vector D.
- One or more filenames each containing a training set.

For generating an initial HMM we take the N equally placed observations (feature vector) from the first training set. Each one is used to train a separate state. After training the states have a mean vector which is of size D. And a variance matrix of size D X D containing all zeros. Then for each of the remaining observations, we find the Euclidean distances between it and the mean vector of the states. We assign an observation to the closest state for training. The state assigned to consecutive observations are tracked to find the transitional probabilities.

Segmental K-means algorithm tries to modify the initial model so as to maximise $P(O, I/\lambda)$, where O are the training sets used for training and I is a state sequence in the given HMM. The maximised (optimal) path for a training set is denoted by I*. Those observations that were assigned to a different state then the one in which they should be present according the optimal path are then moved to the state. This improves $P(O,I*/\lambda)$. The model is evaluated again so with this changed assignments of observations. The above process iteratively till there are no more reassignment needed. The calculation of mean, variance, and transitional probabilities are done as shown before.[8]

Viterbi algorithm is useful identifying the best path that a signal can take in a HMM. Find the best path is a search problem. Viterbi uses dynamic programming to reduce the search space. For the first observation sequence the out probability of a state being the start state. This done by taking a product of initial probability and the observation probability for the state. For every other observation all the states try to find a predecessor such that the probability of the predecessor multiplied by the transition probability from the predecessor to itself is maximised.

### 5. IMPLEMENTATION OF COMMAND INPUT/OUTPUT SYSTEM

The implementation of the command Input / Output system is done by CHMM. Continuous HMM library, which supports vector as observations, has been implemented in the project. The library uses probability distribution functions that are mention in section 3. The system has a model for each word that the system can recognize. The list of words can be considered as language model. While recognizing the system need to know where to locate the model for each word and what word the model corresponds to. This information is stored in a flat file called models in a directory called HMMs. The difference in case of HMM is that the symbol does not uniquely identify a state. The new state is determined by the symbol and the transition probabilities from the current state to a candidate state. The system is trained before a word is recognize as mention in section 4. When a sound is given to the system to recognise, it compares each model with the word and finds out the model that most closely matches with it. The word corresponding to that HMM model is given as the output. [1][2][3]

Training the system for a new word requires the sound files for that word. Feature for the sound file can be extracted using the extract feature command. Train command can be used to train the system. The command needs information such as the number of states that the model should have the size of the feature vector, and the file to be used for training. First argument to the command should be a number indicating the number of states. Second argument is the size of the vector. After this one or more file containing the training data. The output of the Train command is the trained HMM in XML format which should be written to a file and put in the hmms directory. An entry needs to be made in the models file present in the same directory. For recognition sound is recorded using the Raw Recorder program. Then extract the feature to get an mfcc file. Recognise command takes one or more filenames as argument. It tries to recognise the word for each file. We propose to implement this system using the .Net API technology. The use of APIs limit the user's prerequisite of DOT NET knowledge required to develop a working project in DOT NET.

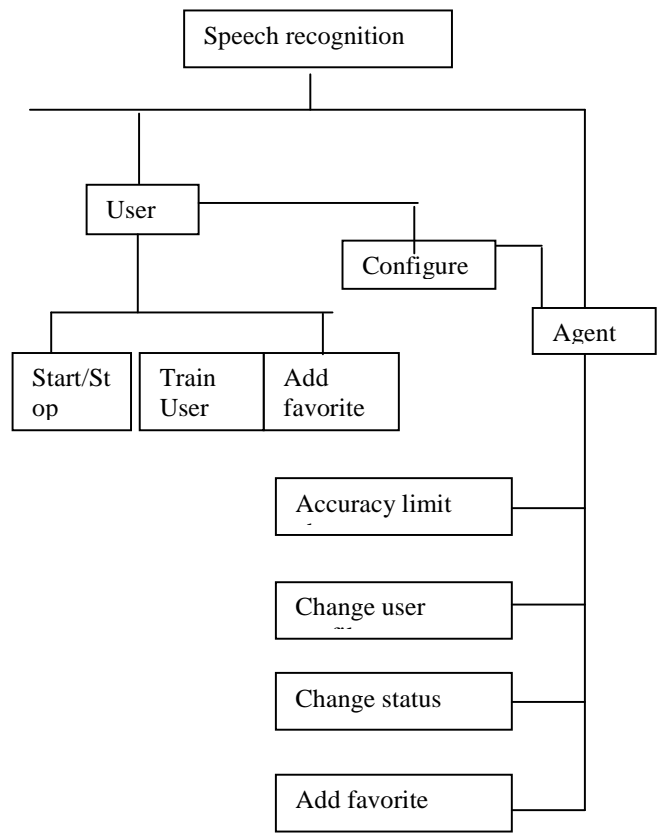

Figure 1 Structural Chart

### 6. EXPERIMENTAL RESULTS

**Training**
To train the system we used 20 users. Each user trained the system . Training of the system was done in a calm and peaceful environment so that recognition accuracy may be more. When user trained the system, a separate profile is created for each user.

## Recognition

Recognition was tried on two kinds of sounds
- Known user: The user whose voice we used for training.
- Unknown user: The user whose voice we not used for training.

The Result of experiment as shown in table:

Table 1: Recognition Result

| Type of user | No of sound | Correct recognition | Incorrect Recognition |
|---|---|---|---|
| Known User | 20 | 18 | 2 |
| Unknown user | 10 | 6 | 0 |

Table 2 shows the command recognition probability and the graphical representation shows in graph 2 on the last page.

Table 2 : Command Recognition Probability Number of Experiments

| Commands | Number of Testing | Recognition Probability |
|---|---|---|
| Activate | 20 | 100% |
| Deactivate | 20 | 100% |
| Welcome | 20 | 100% |
| Word | 20 | 100% |
| Excel | 20 | 100% |
| Save | 20 | 90% |
| Close | 20 | 100% |
| Notepad | 20 | 100% |
| Menu | 20 | 98% |
| Exit | 20 | 100% |
| Escape | 20 | 100% |
| Left | 20 | 100% |
| Right | 20 | 100% |
| Up | 20 | 100% |
| Down | 20 | 100% |
| EnterTheNumricState | 20 | 90% |
| ExitTheNumricState | 20 | 90% |
| EnterAlphabeticState | 20 | 90% |
| ExitTheAlphabeticState | 20 | 90% |
| Plus | 20 | 100% |
| Multiply | 20 | 98% |
| Divide | 20 | 95% |
| Minus | 20 | 100% |
| Star | 20 | 100% |
| Shut down | 20 | 100% |
| PenDriveFormat | 20 | 95% |
| Scanning | 20 | 98% |
| Ok | 20 | 100% |
| Enter | 20 | 100% |
| Run | 20 | 100% |

The system is experimented for 10 number of user. The corresponding recognition percentage result which shown by a graph. This experiment tried to find out how many users is recognition correctly. The outcomes have been exhibited in Table 3 and Graph 3.

Table 3 : Experimental Results

| User | Recognition Percentage |
|---|---|
| A | 85 |
| B | 80 |
| C | 90 |
| D | 95 |
| E | 75 |
| F | 50 |
| G | 65 |
| H | 77 |
| I | 83 |
| J | 98 |

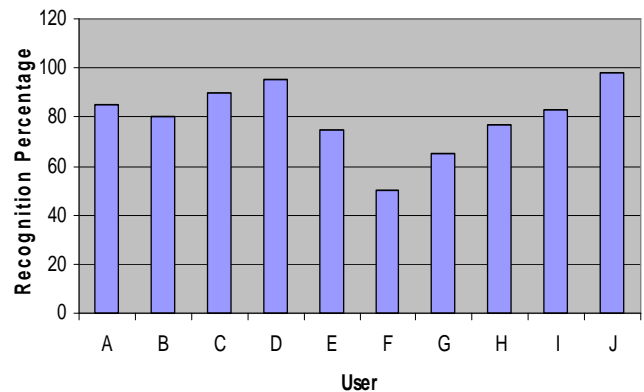

Graph 3. Experimental Results for 10 users

### 7. ACCURACY OF EXPERIMENTS

1. Accuracy of the experiments depends on the training time if the training time of the system has been increased then accuracy automatically increased. Time is directly proportional to accuracy if training time increase then accuracy increased.
2. Accuracy of the system is increased when we trained the system in very peaceful environment and provide the same environment at the time of system used.
3. Accuracy of the system increases when good quality input hardware like microphone is used
4. A graphical representation shows how time plays an important role in accuracy. A table containing number of users and time that user used to train the system is given in Table 4 and Graph 4.

Table 4: Time to accuracy relation

| User | Time to Trained the system (In Hours) | Accuracy in percentage |
|---|---|---|
| Kamlesh | 2 | 70 |
| Lalit | 1 | 45 |
| Kamal | 3 | 95 |
| Neetu | 1.30 | 65 |
| Kanak | 2.30 | 85 |

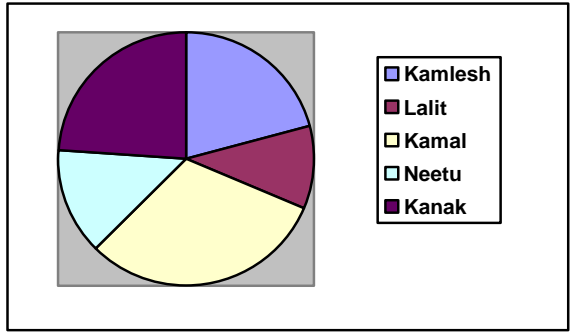

Graph 2: Time to accuracy relation

## 8. CONCLUSIONS AND FUTURE WORK

This paper presents a scheme proposed to control computer systems through voice of different users. The key factor in designing such system is the target audience. For example, physically handicapped people should be able to wear a headset and have their hands and eyes free in order to operate the system. Today, while considering this question, and uses where these technologies will be needed and desire, which would warrant R&D expenditures. There are a number of scenarios where speech recognition is either being delivered, developed for, researched or seriously discussed like computer and video games, precision surgery, domestic applications, wearable computers etc. There are several challenges the system needs to deal with in the future. First, the overall robustness of the system must be improved to facilitate implementation in real life applications involving telephone and computer systems. Second, the system must be able to reject irrelevant speech that does not contain valid words or commands. Third, the recognition process must be developed so that commands can be set in continuous speech. And finally, the voice systems must be able to become viable on low-cost processors. Thus, this will enable the technology to be applied in almost any product.

As with many contemporary technologies, such as the Internet, online payment systems and mobile phone functionality. , development is at least partially driven by the trio of often perceived evils that are "games, gambling and girls (pornography)". Though these applications are outside the educational sphere, it is important to remember that many ICT

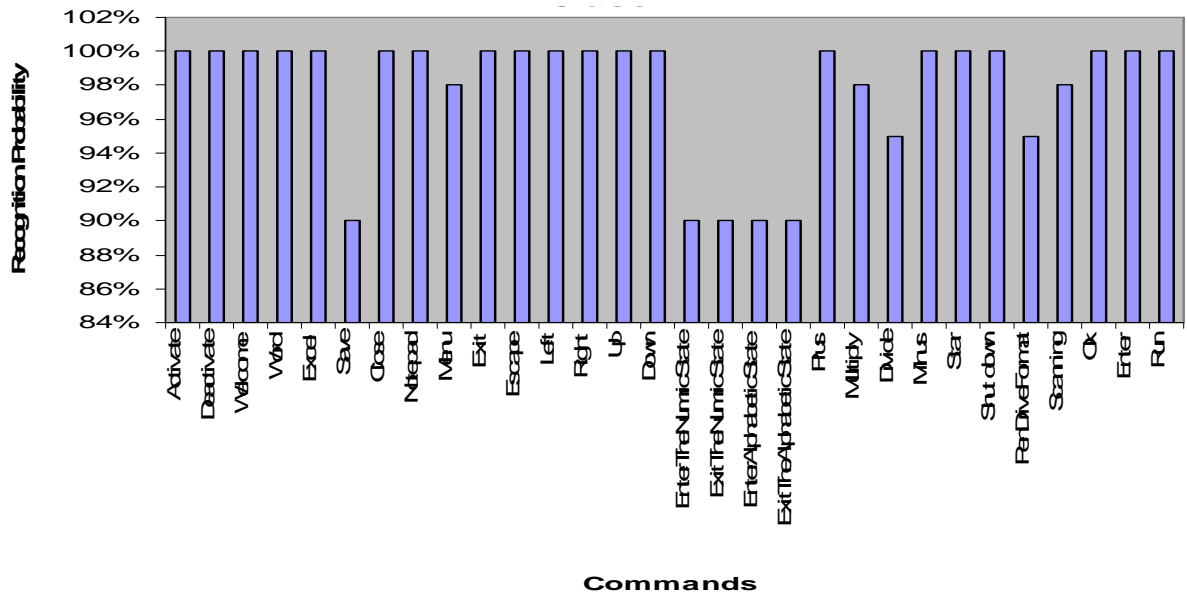

Graph 2 Commands recognition probability for 20 users